\documentclass[conference,10pt, a4paper]{IEEEtran}
\usepackage{xcolor}
\usepackage{mathtools}
\usepackage{epsfig,makeidx,color}
\usepackage{amsmath,amssymb,bbm,enumitem}
\usepackage{cite,graphicx,lipsum}
\usepackage[switch,pagewise]{lineno}
\usepackage{hyperref}
\usepackage[caption=false,font=footnotesize]{subfig}
\usepackage{algorithm}
\usepackage{algorithmic}
\usepackage{acronym}
\usepackage{array}    
\hypersetup{
        colorlinks = true,
        citecolor=red,
}
\usepackage{tikz}

\def\cM{{\cal M}}
\def\cH{{\cal H}}

\def\cG{{\cal G}}

\DeclareMathOperator*{\argmin}{\arg\!\min}
\DeclareMathOperator*{\argmax}{\arg\!\max}

\def\be{ \begin{equation} }
\def\ee{ \end{equation} }
\def\bea{ \begin{eqnarray} }
\def\eea{ \end{eqnarray} }

\def\bx{{\bf x}}
\def\by{{\bf y}}

\def\bn{{\bf n}}

\def\b0{{\bf 0}}

\def\cC{{\cal C}}

\def\cD{{\cal D}}
\def\cF{{\cal F}}
\def\cB{{\cal B}}

\def\cI{{\cal I}}

\def\cM{{\cal M}}
\def\cW{{\cal W}}

\acrodef{TX}{transmitter}
\acrodef{RX}{receiver}
\acrodef{BEC}{bit erasure channel}
\acrodef{MSE}{mean squared error}
\acrodef{LPIPS}{learned perceptual image patch similarity}
\acrodef{AIGC}{AI-generated content}
\acrodef{ML}{machine learning}
\acrodef{NLP}{natural language processing}
\acrodef{LLM}{large language model}
\acrodef{CLIP}{contrastive language-image pre-training}
\acrodef{SemCom}{semantic communication}
\acrodef{SemPA}{semantic packet aggregation}
\acrodef{SemRT}{semantic repeated transmission}
\acrodef{SMART}{semantic packet aggregation and repeated transmission}

\ifCLASSOPTIONonecolumn
\else

\fi

\usepackage{geometry}
\graphicspath{ {./images/} }
\usepackage{graphicx}
\begin{document}


\title{Semantic Packet Aggregation and Repeated Transmission for Text-to-Image Generation}

\author{Seunghun Lee, Jihong Park, Jinho Choi, and Hyuncheol Park}

\maketitle

\begin{abstract} 
Text-based communication is expected to be prevalent in 6G applications such as wireless AI-generated content (AIGC). Motivated by this, this paper addresses the challenges of transmitting text prompts over erasure channels for a text-to-image AIGC task by developing the semantic segmentation and repeated transmission (SMART) algorithm. SMART groups words in text prompts into packets, prioritizing the task-specific significance of semantics within these packets, and optimizes the number of repeated transmissions. Simulation results show that SMART achieves higher similarities in received texts and generated images compared to a character-level packetization baseline, while reducing computing latency by orders of magnitude compared to an exhaustive search baseline.
\end{abstract}

\begin{IEEEkeywords}
Wireless AIGC, Packet Aggregation, Repeated Transmission, Semantic Communication.
\end{IEEEkeywords}

\ifCLASSOPTIONonecolumn
\baselineskip 28pt
\fi
\section{Introduction}


Recent advancements in large language models (LLMs) and generative models rely on vast amounts of training data, primarily consisting of human language. This extensive exposure enables these models to generate human-like text and manage various data modalities through human language \cite{liu24}, positioning texts as a foundational element. In beyond 5G and 6G networks, we anticipate the rise of human language-based wireless applications, such as wireless AI chatbots and \ac{AIGC} applications \cite{Minrui24}. In these applications, mobile devices send text prompts to remote servers hosting LLMs or generative models, which process the prompts and send back the results. 

Building on this trend, this paper tackles the challenges of transmitting text prompts over an erasure channel for a text-to-image \ac{AIGC} task, by developing the \ac{SMART} algorithm, comprising two key components: \ac{SemPA} and \ac{SemRT}. \ac{SemPA} groups words in a text prompt into packets to maximize the expected similarity between the transmitted and received text prompts, or equivalently between the original and generated images. Unlike conventional communication, where all bits hold equal importance and packetization focuses only on optimizing packet length, \ac{SemPA} constructs packets by prioritizing the task-specific significance of meanings or semantics within text packets. Based on the solutions of \ac{SemPA}, \ac{SemRT} determines the number of repeated packet transmissions to overcome erasure channels. To reduce the computation complexity, a supervised machine learning (ML) approach is applied. 

Simulations on the MS-COCO image-caption dataset \cite{MSCOCO14} demonstrate that SMART achieves over 6x higher text cosine similarity and 48.3\% higher image cosine similarity than a baseline that transmits each character separately, while also reducing the computing latency by up to $5\times 10^5$ times compared to an exhaustive search baseline. Note that SMART can be viewed as a form of text-based semantic communication, which has been recently explored but through different perspectives, including LLM-based source and channel coding \cite{Nam2023LanguageOrientedCW}, error correction \cite{Guo22,Guo24}, and multimodal data transmission \cite{cicchetti24}. In \cite{Qin22,Xidong23,Jinsong24}, text similarity is measured using pre-trained language models, though without semantics-aware packtization and transmission strategies.

\begin{figure*}[t]
\centering
\includegraphics[width=\textwidth]{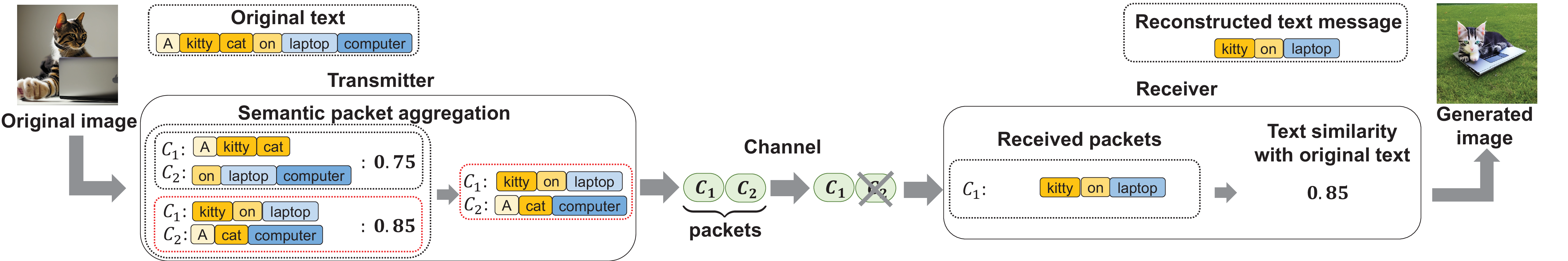}
\vspace{-0.5cm}
\caption{The structure of word group transmission system. When words are grouped based on their semantics, the received packets contain important semantics.}\label{fig:1}
\end{figure*}

\section{Word Group Transmission System}
\subsection{Word-level Message Transmission Scenario}
As shown in Fig.~\ref{fig:1}, for semantic communication, we consider a scenario where the text message of \(K\) words \([w_1, w_2, \cdots, w_K]\), which can represent the original image, is encoded 
and transmitted from a \ac{TX} to a \ac{RX} over an unreliable packet-level channel.
In this scenario, due to a high compression ratio, the reconstruction at RX becomes sensitive to channel errors, i.e., if some words are lost (or erased), RX may not be able to reconstruct an image that conveys the same meaning as the original. 
Thus, to minimize the possibility that important semantics in the message is lost due to packet loss, we divide a text message of words into multiple groups in a way that ensures important semantics are preserved even when some packets are lost, where each group consists of a few words, as illustrated in Fig.~\ref{fig:1}. To this end, forming groups can be optimized for a given channel characteristic, which will be studied in Section~\ref{sec3}. 
For convenience, we denote the groups as \(\cC_1, \cC_2, \cdots, \cC_{|\cG|},\) where \(\cG\) is the set of groups, i.e., \(\{\cC_1, \cC_2, \cdots, \cC_{|\cG|}\}\). Each group consisting of $M$ words is transmitted as a single packet.




\subsection{Channel Model}
We consider a packet-level channel model, since network layer channels are typically modeled as packet-level channels
\cite{Sundararajan11}.
In the channel, we assume that a packet can be erased with a  probability of \(p\). 
Denote by \({\hat{\cC}}\)
the received packet at \ac{RX}, which is given as:
\begin{align}\label{Z}
{\hat{\cC}} = \begin{cases}
{\cC}, &{\text {with probability }} 1 - p, 
\\ \tilde{{\cC}, } &{\text {with probability }} p,
\end{cases}
\end{align}
where \(\tilde{{\cC}}\) is a group in which the \(M\) words are randomly chosen from a pre-defined dictionary \(\cD\) in both \ac{TX} and \ac{RX}. We denote \(\cC\) as the generic notation of word group within \(\cG\).


\subsection{Expected Similarity Score between Different Messages}
We introduce a metric that measures the similarity between semantics in texts based on pre-trained model, such as \ac{CLIP} \cite{CLIP}. 
The cosine similarity score evaluates how closely the feature vectors, to which texts are mapped, align in direction—since the model encodes semantically related texts into similarly oriented vectors, the cosine value naturally reflects the degree of semantic alignment \cite{Guo22}
:
\begin{align}\label{sim_f}
    \phi(x, y) = \frac{g(x)^{\mathrm{T}} g(y)}{||g(x)|| \cdot ||g(y)||},
\end{align}
where \(\phi(\cdot, \cdot)\) is the normalized similarity score function. \(g(\cdot)\) symbolizes the text encoding process of the pre-trained model, transforming inputs into feature vectors.

Since we need to quantify the similarity between the original message and the message decoded from any subset of received groups \(\cC \in \cG\) after packet loss, one of the argument in \(\phi(\cdot, \cdot)\) becomes \({\cW}\). Thus, we omit the second argument and denote \(\phi(\cdot, \cdot)\) as \(\phi(x)\), which is the similarity between \(x\) and \({\cW}\).

\section{Semantic Packet Aggregation}\label{sec3}
In this section, we formulate expected similarity score maximization problem for the unreliable channel. We assign each word in \(\cW\) into one of the groups \(\cC \in \cG\). Since packets are lost randomly, we denote by \( \cH \) the subset of \( \cG \) representing received packets in each loss scenario and aim to maximize the average semantic similarity across all possible \( \cH\) configurations.
Each \(\cC \in \cG\) is transmitted once as a single packet. Therefore, the total number of packet transmissions is \(\frac{K}{M}\).
Letting \(\cF(\cG) =  \cup_{\cC \in \cG}\cC\), the \ac{SemPA} problem for maximizing the expected similarity score can be formulated as:
\refstepcounter{equation}\label{P0}
\begin{flalign*}
\textbf{P0: }&\max \limits_{\cG, M \in {\cM}}\sum_{\cH \subseteq \cG} 
{p}(\cH)\cdot\phi(\cF(\cH))\tag{\theequation a}\label{P0a} &&
\\& {~{\text {s.t.}}}\hspace{0.3pc} \cF({\cG})\!=\!{\cW} ({\text{complete transmission constraint}}),\hspace{-0.5pc}\tag{\theequation b}\label{P0b}&&
\\& \hphantom{~{\textrm {s.t.}}}\hspace{0.3pc}|\cC| = M\hspace{0.3pc}\forall \cC \in \cG ({\text{packet length constraint}}), \tag{\theequation c}\label{P0c}&&
\\& \hphantom{~{\textrm {s.t.}}}\hspace{0.3pc}{\cM} = \{M \in {\mathbb{Z}}^{+} : \frac{K}{M} \in {\mathbb{Z}}^{+}\}, \tag{\theequation d}\label{P0d}
\end{flalign*}
where \({p}(\cH) = (1 - p)^{|\cH|}p^{\frac{K}{M} - |\cH|}\), which is the probability that only \(\cC \in \cH\) are successfully delivered to \ac{RX} and \(\cG\) is partition of \({\cW}\) in \eqref{P0b}.
In \eqref{P0c} and \eqref{P0d}, \(\cM\) is defined as the set of feasible \(|\cC|\).

In \textbf{P0}, no feasible pair in \((\cG,M)\) has different values of \(M\) associated with the same \(\cG\). This allows us to decompose \textbf{P0} into two subproblems of: \textbf{P1} finding the optimal \(\cG^{\star}\) and \textbf{P2} deriving the optimal \(M^{\star} \in \cM\) given \(\cG^{\star}\), which can be solved sequentially.
First, we find the optimal \(\cG^{\star}\) for each \(M \in \cM\) by solving \textbf{P1} below.
\begin{flalign}\label{P1}
\textbf{P1: }&\max \limits_{\cG}\sum_{\cH\subseteq \cG} {p}(\cH)\cdot\phi(\cF(\cH))&&
\\& {~{\text {s.t.}}~}\hspace{0.3pc}\eqref{P0b},\eqref{P0c}.&& \nonumber
\end{flalign}

\begin{figure}
\centering
\includegraphics[width=0.7\columnwidth]{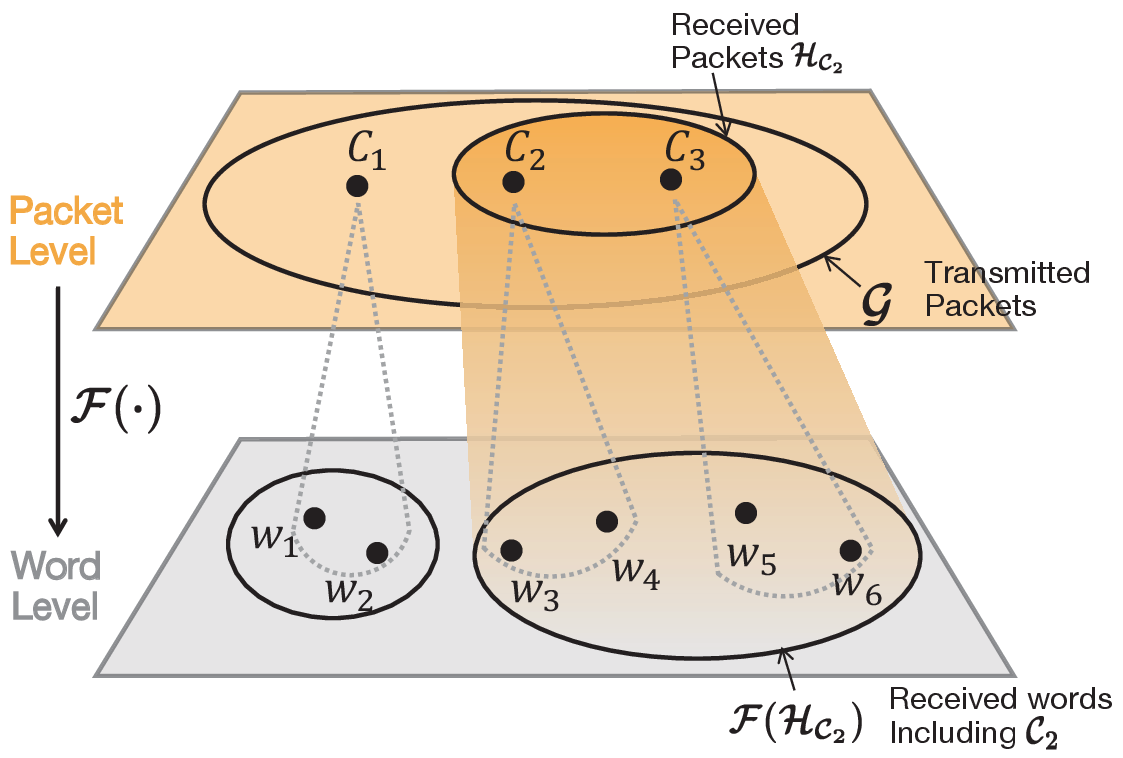}
\caption{Relationships among \(\cG, \cC, \cH_{\cC},\) and \(\cF(\cH_{\cC}) \). In this example, \({\cC_1} = \{w_1, w_2\}, {\cC_2} = \{w_3, w_4\},{\cC_3} = \{w_5, w_6\}\), and \(\cH_{\cC_2} = \{\cC_2, \cC_3\}\).}\label{fig:group_illustration}
\end{figure}

Finding the optimal \(\cG\) in {\textbf{P1}} is computationally challenging. For instance, if exhaustive search is applied, the number of feasible combinations of \((\cG,M)\) that need to be evaluated is \( \sum_{M \in \cM}\frac{K!}{(M!)^{\frac{K}{M}} \times (\frac{K}{M})!}\), which increases combinatorially as \(K\) increases.
To reduce this complexity, we approximate {\textbf{P1}} as a problem of finding \(\cC\) to be included in \(\cG\) and change the optimization variable \(\cG\) to \(\cC\). Such approximation reduces the number of feasible optimization variable into \(\binom{K}{M}\), which is the number of possible \(\cC \subseteq {\cW}\) for each \(M \in \cM\). The appoximation process is illustrated in Fig.~\ref{fig:group_illustration}.
To approximate {\textbf{P1}}, we isolate the term containing \(\cC \in \cG\) from {\textbf{P1}}'s expected similarity score and define the approximated problem as maximizing the isolated term:
\begin{align}\label{G_p}
\sum_{\cH_{\cC} \subseteq \cG } {p}( \cH_{\cC})\cdot\phi(\cF(\cH_{\cC})),
\end{align}
where \(\{\cC\} \subseteq \cH_{\cC} \subseteq \cG\).
To change the optimization variable \(\cG\) to \(\cC\), we approximate \eqref{G_p} by projecting \(\cG\) onto \(\cF(\cG)\), since \(\cF(\cG) = \cW\) is fixed regardless of \(\cG\) from the complete transmission constraint:
\begin{align}
\sum_{\cF({\cH}_{\cC})\subseteq  \cF({\cG})} {q}(\cF({\cH}_{\cC}))\cdot \phi(\cF({\cH}_{\cC})),\label{individual_a}
\end{align}
where \({q}(\cF({\cH}_{\cC})) = (1 - p)^{|\cF({\cH}_{\cC})|}p^{K - |\cF({\cH}_{\cC})|}\).
The accuracy of the approximation from \eqref{G_p} to \eqref{individual_a} can be found in the Table~\ref{table:compare_scores}. The example message is `cat that is wearing yellow hat'. As shown, when comparing the ranks of each feasible \( \cC \) using \eqref{G_p} averaged over possible \(\cG\) and \eqref{individual_a}, there is not a significant difference, where a higher rank corresponds to a higher score.

\begin{table}[t!]
\centering
\caption{Exact and Approximate Expected Similarity Scores with Ranks. A Higher Rank Corresponds to a Higher Score.}
\label{table:compare_scores}
\begin{tabular}{|m{2cm}|m{1.2cm}|m{1cm}||m{1.2cm}|m{1cm}|}
\hline
\textbf{\( \cC \)} & \textbf{Score in} \eqref{G_p} & \textbf{Rank} & \textbf{Score in} \eqref{individual_a} & \textbf{Rank} \\
\hline
(cat, yellow)      & 0.762835 & 1  & 0.736258 & 1 \\
(cat, hat)         & 0.759744 & 2  & 0.734356 & 2 \\
(cat, wearing)     & 0.751852 & 3  & 0.730508 & 3 \\
(cat, that)        & 0.746135 & 4  & 0.722177 & 4 \\
(cat, is)          & 0.743091 & 5  & 0.721482 & 5 \\
(yellow, hat)      & 0.728864 & 6  & 0.711926 & 6 \\
(wearing, yellow)  & 0.716593 & 7  & 0.702556 & 7 \\
(wearing, hat)     & 0.713455 & 8  & 0.700222 & 8 \\
(yellow, that)     & 0.713094 & 9  & 0.695591 & 11 \\
(yellow, is)       & 0.711962 & 10 & 0.698600 & 9 \\
(that, hat)        & 0.710915 & 11 & 0.694204 & 12 \\
(is, hat)          & 0.709316 & 12 & 0.697111 & 10 \\
(wearing, that)    & 0.699603 & 13 & 0.686686 & 13 \\
(wearing, is)      & 0.695966 & 14 & 0.685327 & 14 \\
(is, that)         & 0.693780 & 15 & 0.682176 & 15 \\
\hline
\end{tabular}
\end{table}

In summary, letting \( \psi(\cC) = \sum_{\cF({\cH}_{\cC})\subseteq \cW} {q}(\cF({\cH}_{\cC}))\cdot \phi(\cF({\cH}_{\cC}))\), \(\textbf{P1}\) is recasted as:
\begin{flalign}
\textbf{P1.1: }&\max \limits_{\cC} \psi(\cC)&&
\\& {~{\text {s.t.}}~}\hspace{0.3pc}\eqref{P0c}.&& \nonumber
\end{flalign}
To obtain \(\cG^{\star}\) by solving {\textbf{P1.1}}, top \(\frac{K}{M}\) subsets \(\cC\) that maximize \(\psi(\cC)\) are selected. Thus, \(\cG^{\star}\) is given as:
\begin{align}\label{biased score}
\hspace{-0.4pc}\cG^{\star}\!=\!\{\cC_i\!\mid \cC_i \cap \cC_j\!=\!\emptyset~ \forall \cC_i, \cC_j\!\in\!\cG, \hspace{0.3pc} \cC_i =\hspace{-0.5pc}\argmax_{\cC \subseteq {\cW} \setminus \cup_{k=1}^{i-1}\cC_k}\!\!\psi(\cC_i) \}.
\end{align}

\begin{figure}
\centering
\includegraphics[width=0.7\columnwidth]{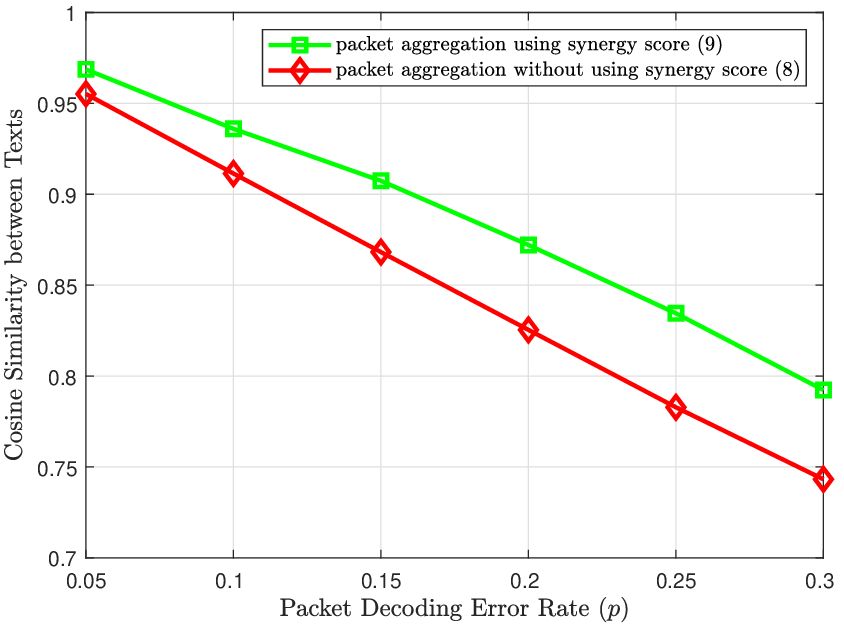}
\caption{Comparison of expected similarity scores across different \(p\) when \(K = 6, M = 2\): grouping methods with and without applying synergy score.}\label{fig:synergy_score_comparison}
\end{figure}

However, when determining sets \( \cC \in \cG\) sequentially in order of \( \psi(\cC)\), most important words in \(\cW\) are more likely to be grouped. If such a group fails to be transmitted successfully as a packet, the similarity score significantly decreases. To prevent such tendency, we introduce a baseline \(b(\cC)\) defined by the scores obtained when transmitting each word in \( \cC\)  as an individual packet and select groups in ascending order of \((b(\cC)-\psi(\cC))\):
\begin{align}\label{synergy score}
\cG^{\star}\!=\{\cC_i\!\mid &\cC_i \cap \cC_j\!=\!\emptyset \hspace{0.3pc} \forall \cC_i, \cC_j \in \cG, \nonumber \\ 
&\cC_i = \argmin_{\cC_i \subseteq {\cW} \setminus \cup_{k=1}^{i-1}\cC_k}\!\!\left(b(\cC_i) - \psi(\cC_i)\right)\},
\end{align}
where \(b(\cC) = \sum_{\substack{ {\cD_{\cC}} \subseteq {\cW}}} r( {\cD_{\cC}})\cdot \phi({\cD_{\cC}})\) is the baseline, \(\cD_{\cC} \cap \cC \neq \emptyset\), and \(r({\cD}_{\cC}) = (1 - p)^{|\cD_{\cC}|}p^{K - |\cD_{\cC}|}\). 
Fig.~\ref{fig:synergy_score_comparison} illustrates the effectiveness of \eqref{synergy score} by comparing it to \eqref{biased score} across different \(p\) values for \(K = 6, M = 2\), using the sample message `cat that looks like a tiger'.

As \(\cG^{\star}\) is determined for \(M \in \cM\) by solving \textbf{P1.1}, we construct \(\textbf{P2}\) as below:
\begin{flalign}
\textbf{P2: }&\max \limits_{M \in \cM} \sum_{\cH\subseteq \cG^{\star}} 
{p}(\cH)\cdot\phi(\cF(\cH)) \label{P2a} &&
\\& {~{\text {s.t.}}~}\hspace{0.3pc}\eqref{P0c},\eqref{P0d}.&& \nonumber
\end{flalign}


The optimal \(M^{\star}\) is selected by searching \(M\) that corresponds to the highest score achieved by its respective \(\cG^{\star}\):
\begin{align}\label{P2_sol}
    M^{\star} = \argmax_{M \in \mathcal{M}} \sum_{\cH\subseteq \cG^{\star}} 
{p}(\cH)\cdot\phi(\cF(\cH)).
\end{align}

\section{Semantic Repeated Transmission}

In \(\textbf{P1}\) and \(\textbf{P2}\), the optimization focused on partitioning \(\cW\) into \(\cG\). To attain diversity in channel, we extend the problem to scenario that allows adjusting the number of transmissions. At the same time, we generalize \textbf{P0} by relaxing the condition of transmitting all \(\cC \in \cG\), allowing only a subset of groups to be transmitted. Specifically, each word group \(\cC\) can be transmitted \( n_i \) times and the number of total packet transmissions as \(\frac{K}{M}\) which is same as \textbf{P0}, \(0 \leq n_i \leq \frac{K}{M}\) and \(\frac{K}{M} = \sum_{i=1}^{|\cG|}n_i\). For brevity, we denote \(\frac{K}{M}\) as \(N\). Unlike {\textbf{P0}}, \(|\cG|\) could exceed \(N\), since some \(n_i\) could be zero. The \ac{SemRT} problem is given as:
\refstepcounter{equation}\label{P3}
\begin{flalign*}
\textbf{P3: }&\max \limits_{\cG,M,\bn}\sum_{{\cH}\subseteq {\cG}} 
{p}(\cH)\cdot \phi(\cF({\cH}))\tag{\theequation a}\label{P3a} &&
\\& {~{\textrm {s.t.}}~}M\cdot\sum_{i=1}^{|\cG|}n_i \leq K\tag{\theequation b}\label{P3b}&&
\\& \hphantom{~{\text {s.t.}}~}\hspace{0.3pc}\eqref{P0b},\eqref{P0c},\eqref{P0d},&& \nonumber
\end{flalign*}
where \({p}({\cH}) = \prod_{\substack{i \in {\cI}(\cH), \\ j \in {\cI}(\cG \setminus \cH)}}(1 - p^{n_i}) p^{n_j}\) is the probability that only \(\cC \in \cH\) are successfully delivered to \ac{RX} and \(\cI(\cH)\) is the index set of the elements in \(\cH \).

In \eqref{P3b}, \(L \triangleq M \cdot\sum_{i=1}^{|\cG|}n_i \leq K\) is the word transmission limits, which represents the communication latency constraint. 
As each \(\cC\in\cG\) is encoded word by word and \(|\cC| = M\), \(\forall \cC \in \cG\) are encoded into a fixed-length bits. Those encoded bits are proportional to the number of words encoded.
In this context, we apply the word transmission count constraint using \(L\) for the latency constraint. Note that {\textbf{P0}} and {\textbf{P3}} satisfies the same latency constraint, since the number of words transmitted in {\textbf{P0}} is \(M\cdot \frac{K}{M} = K\).

To solve \textbf{P3}, we approximate the problem in the same manner as the approximation from \textbf{P1} to \textbf{P1.1}.
When \(|\cG| \leq N\), \eqref{synergy score} can be applied to find \(\cG^{\star}\). When \(|\cG| > N\), \(\cG\) no longer requires disjointness of \(\cC\), resulting in a higher number of feasible \(\cG\) combinations compared to \textbf{P1}. Furthermore, the approximation of \eqref{G_p} to \eqref{individual_a} cannot be directly applied, since \(\cC \in \cG\) are not disjoint. Finding the optimal \(\cG^{\star}\) among whole feasible \(\cG\) is left for future work. Here, we initially select the first \(N\) groups to be disjoint to satisfy complete transmission constraint and additional \(\left(|\cG| - N \right)\) groups are selected in descending order of their synergy scores. In addition, we derive \(M\) in the same manner in {\textbf{P2}}, assuming that \(\cG\) consists of first selected \(N\) groups. Simulations in \ref{sec5} show the effectiveness of our design. In summary, \(\cG^{\star}\) is given as:
\begin{align}\label{revised synergy score}
{\cG}^{\star}\!=\{\cC_i\!\mid &\cC_i \cap \cC_j\!=\!\emptyset \hspace{0.3pc} \forall 0 \leq i<j \leq N, \nonumber \\ 
&\cC_i = \argmin_{\cC_i \subseteq {\cW} \setminus \cup_{k=1}^{i-1}\cC_k}\!\!\left(b(\cC_i) - \psi(\cC_i)\right)\}.
\end{align}

After \({\cG^{\star}},M^{\star}\) are determined, \textbf{P3} is casted as:
\refstepcounter{equation}\label{P3.1}
\begin{flalign*}
\textbf{P3.1: }&\max \limits_{\bn} \sum_{{\cH}\subseteq {\cG^{\star}}} 
{p}(\cH)\cdot \phi(\cF({\cH})) \tag{\theequation a}\label{P3.1a} &&
\\& ~{\textrm {s.t.}}\hspace{0.3pc} \sum_{i=1}^{|\cG^{\star}|}n_i \leq N, \tag{\theequation b}&&
\\& \hphantom{~{\text {s.t.}}~}\hspace{0.3pc}\eqref{P3b}.&& \nonumber
\end{flalign*}
The optimal \(\bn\) in \(\textbf{P3.1}\) can be found by exhaustive search. To reduce the complexity of finding optimal \(\bn\) for every new message, we propose the supervised \ac{ML} method.
The neural network \(h_{\theta}(\cdot)\) is trained to derive \(\bn\) which approximates the optimal \(\bn^{\star}\) obtained by exhaustive search for the given \(\cG^{\star}\).
In particular, \(\phi(\cF(\cH))\) of all possible \(\cH \in {\cG^{\star}}\) are input of the neural network. The output is \(\hat{\by}\), a \(|\cG^{\star}|\)-dimensional vector representing the initial predicted values for \(\bn^{\star}\). 
The neural network is defined as:
\begin{align}
\hspace{-0.8pc}h_{\theta} : \left[ \phi(\cF(\cC_1)), \phi(\cF(\cC_2)), \cdots, \phi(\cF(\cG^{\star}))\right] \rightarrow [0, N]^{|\cG^{\star}|},
\end{align}
where \([0, N]\) represents the set of integers from \(0\) to \(N\) and \(\theta \) is the trainable parameter of the neural network.
Hence, the dimensions of the input and the output are \(2^{|\cG^{\star}|} - 1\) and \(|\cG^{\star}|\), respectively. 

The training data are generated by conducting an exhaustive search on each \(k\)th training example \(\bx^{(k)}\) to obtain the optimal \(\bn^{(k)}\) and using these results as labels \(\by^{(k)}\).
The dataset to train \(h_{\theta}\) is denoted as \(\mathcal{B}\). For each \(k\)th training example, \({\cC}_{i}^{(k)} \in {\cG}^{(k)}\) are obtained using the proposed grouping method.
Concretely, the \(k\)th data, \( \mathbf{b}_k \in \cB \) is denoted as below:
\begin{align}
&\mathbf{b}^{(k)}\!=\![{\mathbf{x}}^{(k)}, {\by}^{(k)}], \; \mathcal{B}\!=\!\{\mathbf{b}^{(k)}\}_{k=1}^{|\cB|},\\
&{\mathbf{x}}^{(k)}\!=\!\left[ \phi(\cF(\cC_1^{(k)})), \phi(\cF(\cC_2^{(k)})), \cdots, \phi(\cF(\cG^{(k)}))\right],\\
&{\by}^{(k)}\!=\!\bn^{(k)}\!=\![n_{1}^{(k)}, n_{2}^{(k)},\!\cdots\!, n_{|\cG^{(k)}|}^{(k)}].
\end{align}

The \(\mathbf{x}^{(k)}\) contains the similarity scores of the combinations of \({\cC}_{i}^{(k)} \in {\cG}^{(k)}\).
Likewise, only similarity scores are considered as input of the neural network during training, instead of \({{\cG}}^{(k)}\) itself.
In terms of the neural network structure, \(h_{\theta}\) consists of fully connected layers (FCN), batch normalization layers (BN), and dropout layers. Each input \(\{ \bx^{(b)} \}_{b \in \mathcal{J}}\), where \( \mathcal{J} \) is the index set of the batch, is processed through the following sequence: FCN, BN, ReLU activation, and dropout. 
This sequence is repeated for the two hidden layers in the network:
\begin{align}
\hat{\by}^{(k)} = h_{\theta}({\bx}^{(k)}) = [y_{1}^{(k)}, y_{2}^{(k)}, \cdots, y_{|\cG^{\star}|}^{(k)}].
\end{align}

In order to obtain a neural network that predicts the optimal transmission counts \(\bn\), we minimize the following \ac{MSE} loss function that measures the difference between \(\by\) and \(\hat{\by}\):
\begin{align}\label{MSE_loss}
\mathcal{L}(h_{\theta}(\mathbf{x}), \mathbf{y}) = \frac{1}{|\mathcal{J}|} \sum_{k \in \mathcal{J}} \| \hat{\by}^{(k)} - \by^{(k)} \|^2,
\end{align}
where \(\| \cdot \|^2\) denotes the squared 2-norm.
The \ac{MSE} loss function is adopted, which is suitable for regression tasks where the goal is to predict \(n_i \in [0,1,\cdots,N]\).
The update rule for the parameters \(\theta\) is given by:
\begin{align}\label{theta_update}
\theta \leftarrow \theta - \eta \nabla_{\theta} \mathcal{L}(h_{\theta}(\mathbf{x}), \mathbf{y}),
\end{align}
where \(\eta\) is the learning rate.

After training, the model outputs the predicted counts \( \hat{\by} = [y_1, y_2, \cdots, y_{|\cG^{\star}|}]\) for the given \(\cG^{\star}\). \( \hat{\by} \) are adjusted to ensure their sum equals \(N \). Concretely, they are iteratively adjusted based on the differences between the predictions and the rounded values. If the difference is positive, the count with the largest positive difference is incremented:
\begin{align}
i^{\star} &= \arg\max_i (y_i - \lfloor y_i \rfloor), \\
y_{i^{\star}} &\leftarrow y_{i^{\star}} + 1.
\end{align}
If the difference is negative, the count with the largest negative difference is decremented:
\begin{align}
i^{\star} &= \arg\min_i (y_i - \lfloor y_i \rfloor), \\
y_{i^{\star}} &\leftarrow y_{i^{\star}} - 1.
\end{align}
Finally, the adjusted transmission counts \(\hat{\by} \) are returned. 

\section{Simulation Results}\label{sec5}
We evaluate the proposed \ac{SMART} by comparing them with baseline methods, using the MS-COCO image-caption dataset \cite{MSCOCO14}. For our simulations, we sample 100 sentences corresponding to 100 different images from the dataset, with each sentence containing 6 words. We assume that \([w_1, w_2, \cdots, w_K]\) are included in the pre-defined dictionary \( \cD = \{v_1, v_2, \cdots, v_{|\cD|}\}\) in both \ac{TX} and \ac{RX}. The words in each packet are encoded to indices in the dictionary, which needs \(\lceil \log_2 |\cD| \rceil\) bits for each word. For dictionary construction, we generate a fixed-size dictionary comprising 200,000 words extracted from the captions in the dataset. The packet decoding error rate \(p\) ranges from \(0.05\) to \(0.3\) for the packet transmission. For the extended problem, we evaluate the \ac{SMART} where \ac{SemRT} is optimized by supervised ML.
For supervised ML method, the neural network is trained over 100 epochs using the Adam optimizer with a learning rate of 0.001 on a dataset of 3,000 examples. \(|\cG|\) is set as 8.

The performance is measured by computing cosine similarity scores between texts using a pre-trained \ac{CLIP} model. 
Furthermore, to consider the scenarios for performing text-based tasks, we generate \ac{AIGC} from the decoded message at the \ac{RX} using the Stable Diffusion v2.1 model \cite{rombach2022} and measure the similarity to image generated from the original message, which is expressed as:
\begin{align}
\zeta = \frac{g(S(\cF({\hat{\cC}}_i))^{\mathrm{T}} g(S({\cW}))}{||g(S(\cF({\hat{\cC}}_i))|| \cdot ||g(S({\cW}))||},
\end{align}
where \(S(\cdot)\) is the image generation function using the input message and the Stable Diffusion model.

The performance of \ac{SMART} is compared to character-level transmission. Each character is encoded into 8 bits using ASCII encoding, with a total bit transmission limit \(B\). For the proposed method, \(\lceil\log_2 |\cD| \rceil\) bits are used to represent each word, and the total bit constraint is given by \(B = L\lceil\log_2 |\cD|\rceil\). For simplicity, we use \(L\) to represent the bit constraint, since \(|\cD|\) is fixed throughout the simulations. 
Character-level transmission often requires more bits than the proposed scheme, which only needs \(L=6\) to transmit all \(\cC \in \cG\). For character-level transmission, we consider two cases: \(L=12\), where \(B = 12\lceil\log_2 |\cD|\rceil\), and \(L = \infty\), where all characters in the message are transmitted once without constraint. For the proposed scheme, \(L=6\) is sufficient to transmit all words in the message, but we also consider \(L=12\), allowing each \(\cC \in \cG\) to be transmitted twice for diversity gain.

\begin{figure}
     \centering
     \subfloat[Comparison of cosine similarity across different packet transmission schemes.]{%
         \includegraphics[width=0.7\columnwidth]{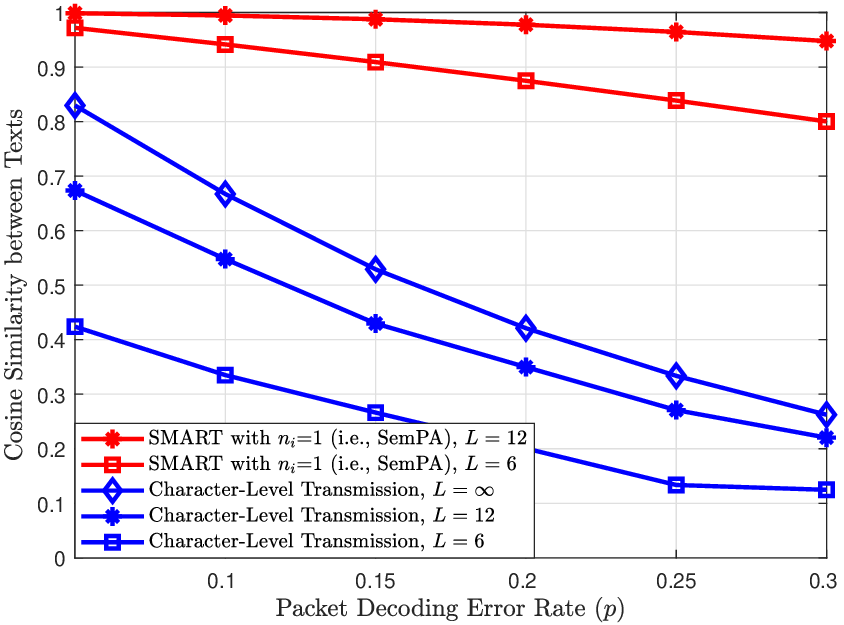}
         \label{fig:plot_textCS_char_a}
     }
     \hfill
     \subfloat[Comparison of cosine similarity across different repeated transmission schemes.]{%
         \includegraphics[width=0.7\columnwidth]{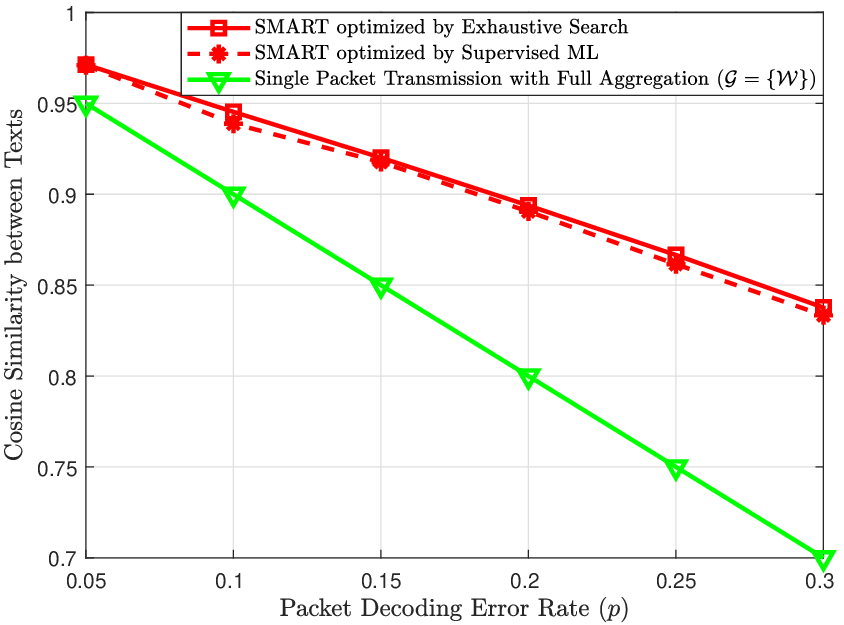}
         \label{fig:plot_textCS_ML}
     }
     \caption{Comparison of cosine similarity between original and decoded messages using different packet transmission methods.} 
     \label{fig:plot_textCS_char}
\end{figure}

\subsection{Text-level Similarity}
We examine the average cosine similarity over different packet transmission methods, which are compared in Fig.~\ref{fig:plot_textCS_char}.
As shown in Fig.~\ref{fig:plot_textCS_char_a}, \ac{SMART} achieves a similarity score more than six times higher than that of character-level transmission at \(p=0.3, L=6\). This implies that, in the case of character-level transmission, character-level errors cause significant distortion of the overall message meaning. 

\begin{figure*}[t]
    \centering

    \begin{minipage}{0.45\textwidth}
        \centering
        \subfloat{%
            \includegraphics[width=0.78\linewidth]{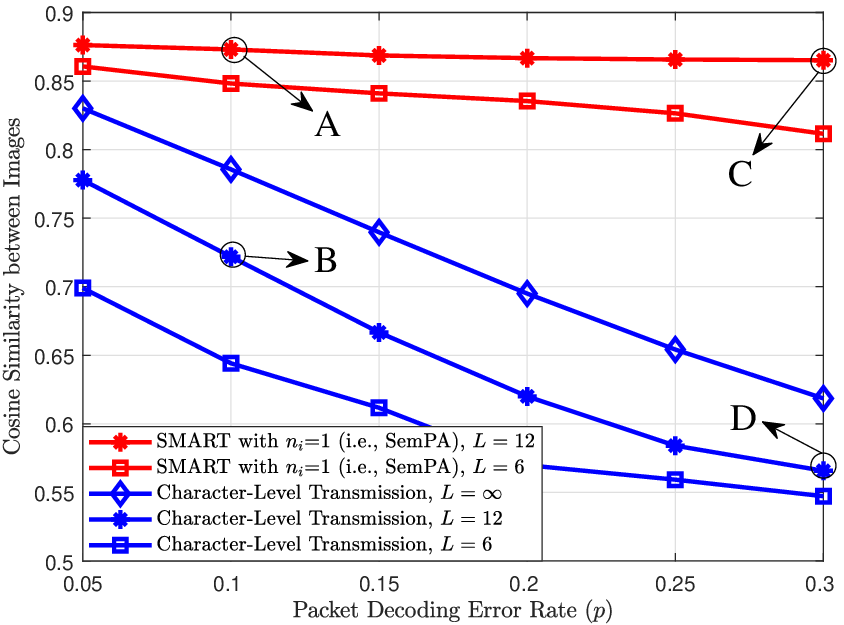}
            \label{fig:plot_textCS_char_b}
        }
    \end{minipage}
    \hspace{1pc}
    \begin{minipage}{0.01\textwidth}
     \vspace{1em} 
     \rotatebox{90}{\textbf{Character-level}\hspace{5pc}\textbf{SMART}}
    \end{minipage}
    \hspace{0.5em} 
    \captionsetup[subfloat]{labelformat=empty}
    \begin{minipage}{0.45\textwidth}
        \centering
        \renewcommand{\thesubfigure}{}
        \begin{minipage}{0.28\textwidth}
            \centering
            {\(p = 0.1\)}
            \subfloat[A: cat that looks like a tiger]{%
                \includegraphics[width=\linewidth]{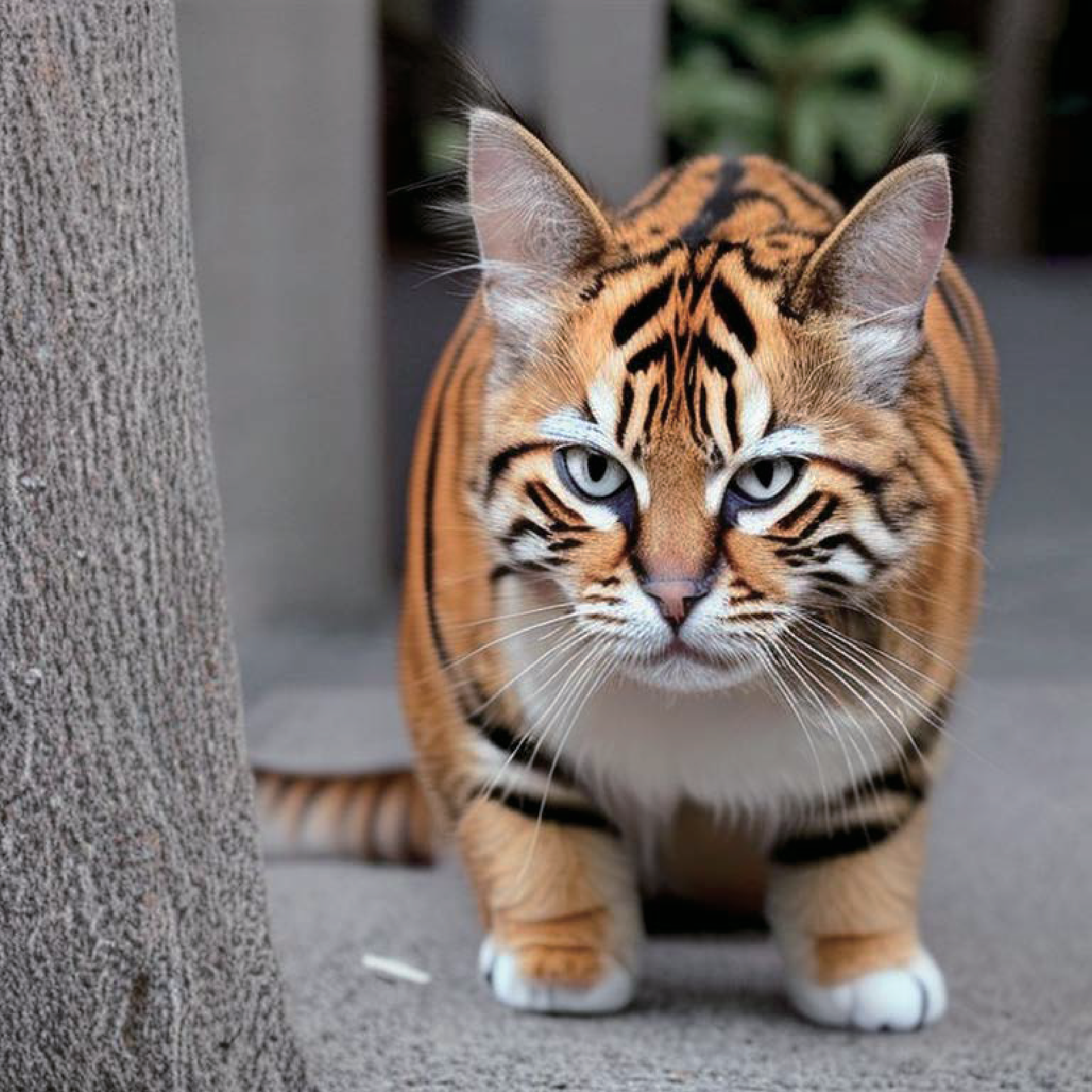}
                \label{fig:wgt_p01}
            }
            \vspace{0.5em} 
            \begin{tikzpicture}
                \draw[dashed, thick] (-1.3,0) -- (1.3,0); 
            \end{tikzpicture}
            \subfloat[B: cat that looks like Wiger]{%
                \includegraphics[width=\linewidth]{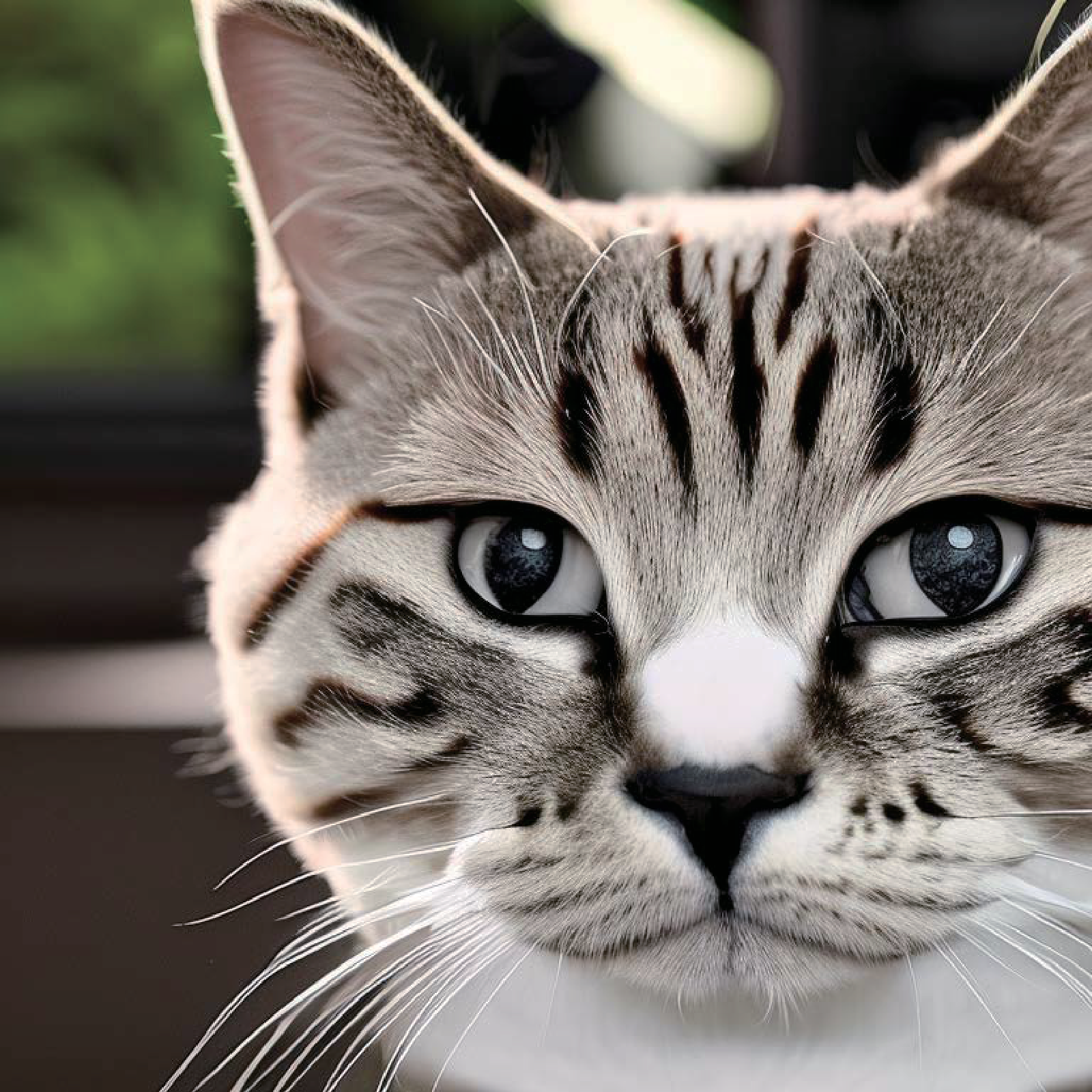}
                \label{fig:char_p01}
            }
        \end{minipage}
        \hspace{0.3pc}
        \begin{minipage}{0.01\textwidth}
            \centering
            \begin{tikzpicture}
                \draw[dashed, thick] (0, 0) -- (0, -7); 
            \end{tikzpicture}
        \end{minipage}
        \begin{minipage}{0.28\textwidth}
            \centering
            {\(p = 0.3\)}
            \subfloat[C: cat that like tiger]{%
                \includegraphics[width=\linewidth]{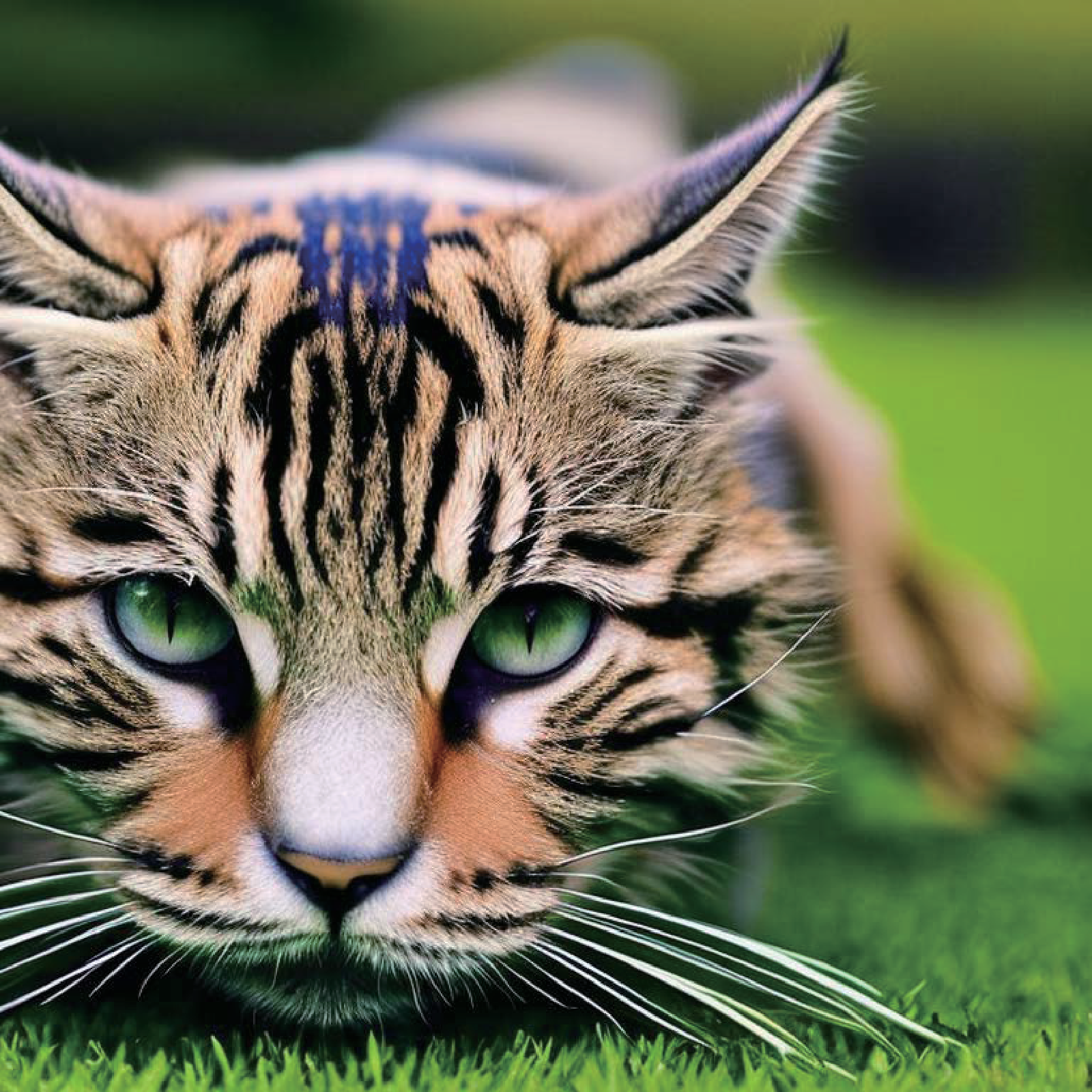}
                \label{fig:wgt_p03}
            }
            \vspace{1.3em} 
        
            \begin{tikzpicture}
                \draw[dashed, thick] (-1.3,0) -- (1.2,0); 
            \end{tikzpicture}
            \hspace{0.3pc}
            \subfloat[D: xam thal YookD lCke tiger]{%
                \includegraphics[width=\linewidth]{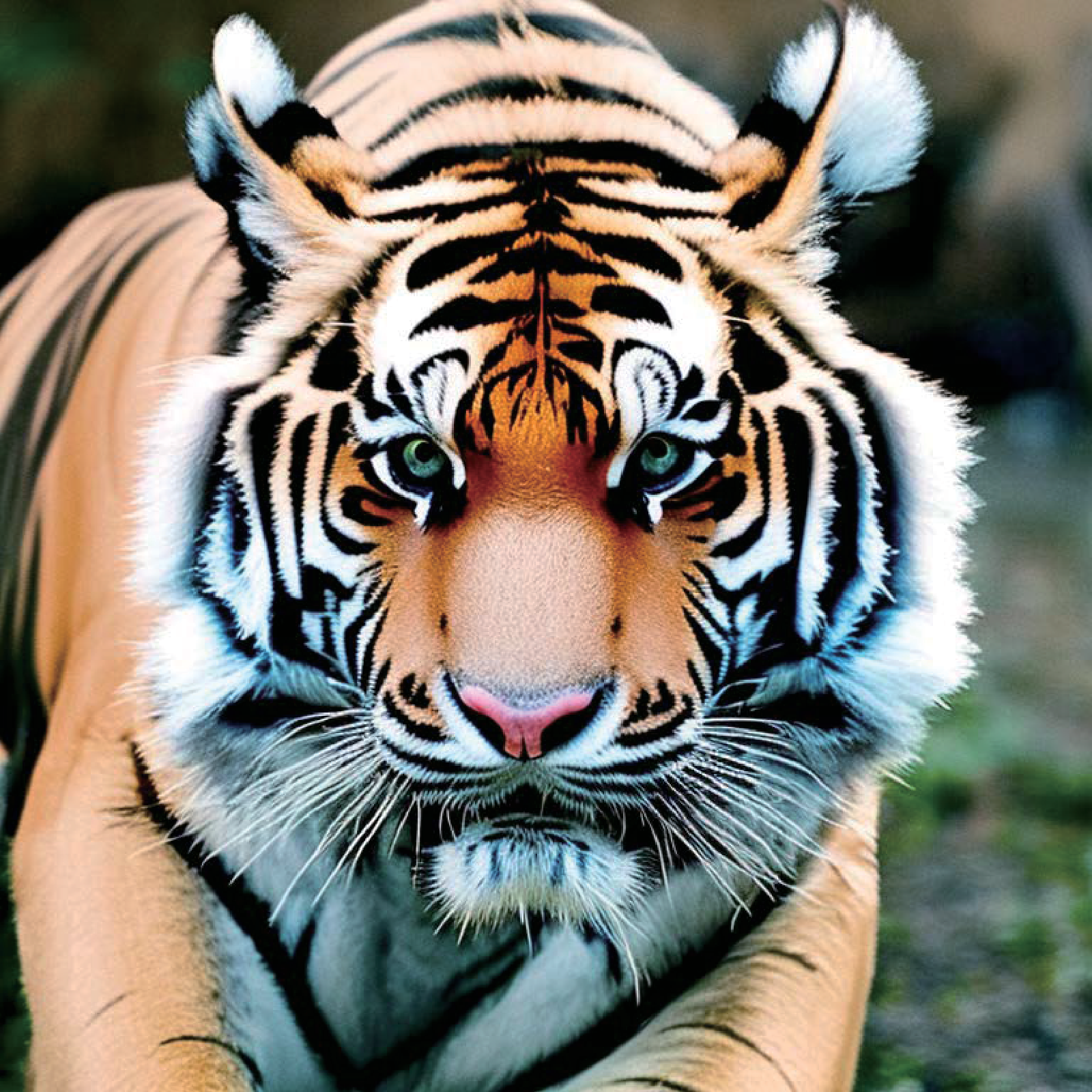}
                \label{fig:char_p03}
            }
        \end{minipage}
        \hspace{0.7pc}
        \begin{minipage}{0.28\textwidth}
            \centering
            \subfloat[Ground truth: cat that looks like a tiger]{%
                \includegraphics[width=\linewidth]{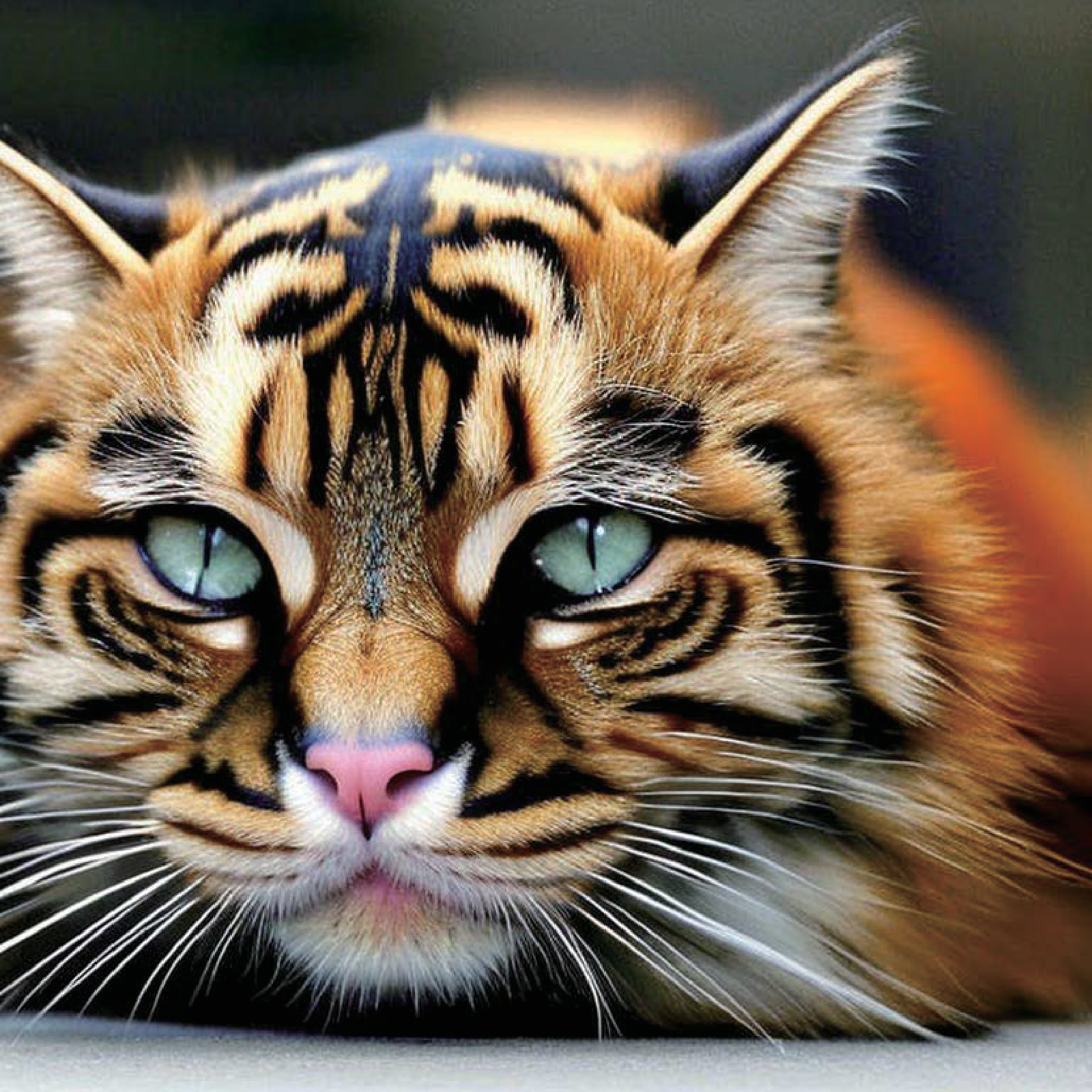}
                \label{fig:ground_truth}
            }
        \end{minipage}

    \end{minipage}

    \caption{Comparison of cosine similarity (left column) between images generated from received messages and original messages across different packet transmission schemes. The example figures in the right column illustrate transmission results, with the first and second columns showing results for \(p=0.1\) and \(p=0.3\) for both \ac{SMART} and character-level transmission, respectively, while the third column shows the ground truth.}
    \label{fig:combined_comparison}
\end{figure*}
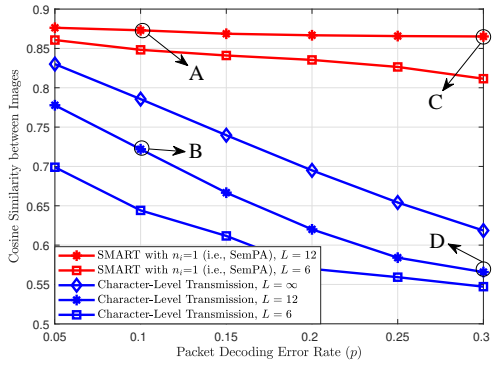
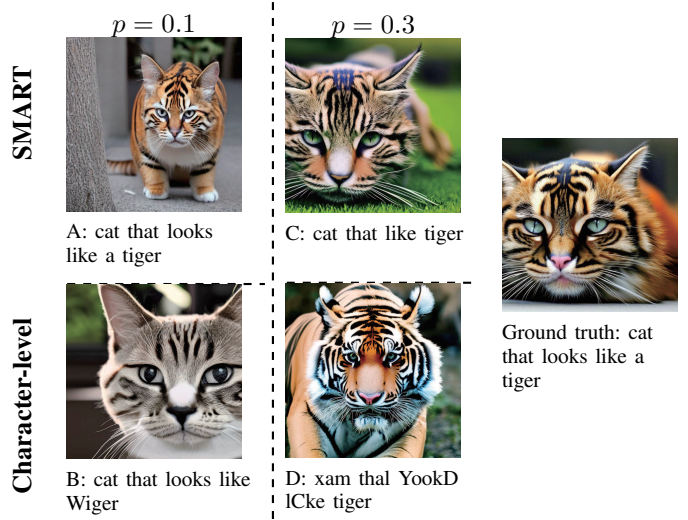

\begin{table}
\centering 
\caption{Average Elapsed Time Comparison for Exhaustive Search vs. Supervised ML for \ac{SMART} (in milliseconds) when \(K = 12, L=12\).} \label{table:complexity_comparison}
\begin{tabular}{|c|c|c|} 
\hline  \(M\) & \textbf{Exhaustive Search (ms)} & \textbf{ML-based Method (ms)} \\ \hline 
1 & \(4.30 \times 10^{5}\) & 0.87 \\ \hline
2 & \(1.36 \times 10^{3}\) & 0.86 \\ \hline
3 & 81.2 & 0.88 \\ \hline
4 & 16.4  & 0.87 \\ \hline
6 & 3.1 & 0.87 \\ \hline
\end{tabular} 
\end{table}

Furthermore, we evaluate the performance of \ac{SMART} in \textbf{P3}. We compare the proposed method to the method that all words in \(\cW\) are fully aggregated as single packet, i.e., \(\cG = \{\cW\}\).
As illustrated in Fig.~\ref{fig:plot_textCS_ML}, optimizing \(\bn\) using supervised \ac{ML} results in a \(19\%\) higher performance compared to \(\cG = \{\cW\}\) when \(L=6, p = 0.3 \). 
The \ac{ML}-based method achieves performance nearly identical to that of exhaustive search for determining optimal \(\bn\), attaining a \(99.4\%\) similarity at \( p = 0.3 \) compared to exhaustive search.

From Table~\ref{table:complexity_comparison}, when \(K\) and \(|\cG|\) increases to \(12\), the exhaustive search method takes an average of \(1.36\,\)s to compute for \(M = 2\), whereas the supervised ML-based method requires only \(0.86\,\)ms. As \(K\) and \(L\) increase, the computational time for exhaustive search for low \(M\) grows exponentially due to the combinatorial explosion, while the ML-based method maintains a nearly constant computational time. Thus, it can be seen that the proposed \ac{ML}-based method is a practical approach that can achieve near-optimal performance while reducing computational complexity.


\subsection{Generated Image-level Similarity}
To evaluate the performance of text-to-image \ac{AIGC} task, Fig.~\ref{fig:combined_comparison} compares the cosine similarity between images generated from transmitted messages and those generated from the original messages.
Despite the challenging task of reconstructing an original-like image from image information transmitted in a text message, the proposed method achieves the highest similarity across all \( p \) ranges, similar to the text-level similarity results shown in Fig.~\ref{fig:plot_textCS_char_a}, demonstrating a \(48.3\%\) higher similarity compared to character-level transmission at \( p = 0.3, L=6 \). 

From the example of generated messages from different packet transmission schemes, \ac{SMART} successfully conveys the message successfully for \(p = 0.3\), delivering the essential information that the cat looks like a tiger. In contrast, character-level transmission results in significant distortions, altering the original sentence to `A xam thal YookD lCke tiger'. This example shows the effectiveness in transmitting the meaning of the original sentence compared to character-level transmission.

\section{Conclusions}
In this paper, we have studied an optimal word grouping strategy for text-based communication that improves the preservation of important semantics, measured by the cosine similarity over unreliable channels. In addition, by leveraging \ac{ML}-based optimization, our approach effectively optimizes the packet transmission count and mitigates transmission errors. Future research will explore grouping methods based on tokenization, as commonly used in LLMs, to further improve the preservation of key meanings within sentences in dynamic channel environments.

\section{Acknowledgement}
This work was partly supported by the Institute of Information \& Communications Technology Planning \& Evaluation(IITP)-ITRC(Information Technology Research Center) grant funded by the Korea government(MSIT)(IITP-2025-RS-2023-00259991, 50\%) and Institute of Information \& communications Technology Planning \& Evaluation (IITP) under 6G·Cloud Research and Education Open Hub grant funded by the Korea government(MSIT)(IITP-2025-RS-2024-00428780, 50\%)

\bibliographystyle{IEEEtran}
\bibliography{si}
\end{document}